# Hexagonal Ti$_2$B$_2$ monolayer: A Promising Anode Material Offering High Rate Capability for Li-Ion and Na-Ion Batteries

Tao Bo[a,b], Peng-Fei Liu[a,b], Juping Xu[a,b], Junrong Zhang[a,b], Fangwei Wang[a,b,c], Bao-Tian Wang[a,b,*]

[a] Institute of High Energy Physics, Chinese Academy of Science (CAS), Beijing 100049, China

[b] Dongguan Neutron Science Center, Dongguan 523803, China

[c] Beijing National Laboratory for Condensed Matter Physics, Institute of Physics, Chinese Academy of Sciences, Beijing 100080, China

E-mail: wangbt@ihep.ac.cn

**ABSTRACT**:

Combining first-principles density functional method and crystal structure prediction techniques, we report a series of hexagonal two-dimensional (2D) transition metal borides (TMBs) including Sc$_2$B$_2$, Ti$_2$B$_2$, V$_2$B$_2$, Cr$_2$B$_2$, Y$_2$B$_2$, Zr$_2$B$_2$, and Mo$_2$B$_2$. Their dynamic and thermal stabilities are testified by phonon and molecular dynamics simulations. We investigate the potential of 2D Ti$_2$B$_2$ monolayer as the anode material for Li-ion batteries (LIBs) and Na-ion batteries (NIBs). The Ti$_2$B$_2$ monolayer possesses high theoretical specific capacities of 456 and 1027 mAhg$^{-1}$ for Li and Na, respectively. The very high Li/Na diffusivity with ultralow energy barrier of 0.017/0.008 eV indicates an excellent charge-discharge capability. In addition, the good electronic conductivity during the whole lithiation process is found by electronic structure calculations. The very small change in volume after the adsorption of one, two, and three layers of Li and Na ions indicates that the Ti$_2$B$_2$ monolayer is robust. These results highlight the suitability of Ti$_2$B$_2$ monolayer as well as the other 2D TMBs as excellent anode materials for both LIBs and NIBs.

# 1. Introduction

Research on advanced energy storage technology is significant for the development of modern society [1]. Rechargeable lithium-ion batteries (LIBs), one of the widely studied clean energy-storage technologies, have attracted increasing attention due to the combination of outstanding reversible capacity, high power density, superior energy efficiency, long cycle life, and portability [2-9]. Currently, LIBs are widely applied in portable electronics, electric vehicles, and electricity grid systems [10, 11]. In addition to LIBs, an alternative option is the rechargeable Na-ion batteries (NIBs), which have been found promising and can be a good candidate to replace LIBs in the future because sodium is more abundant and cheaper than lithium [12-16]. The capability of LIBs and NIBs is highly dependent on the performances of their electrode materials [17, 18]. Nowadays, graphite is commercially used as anode material for LIBs because of its high Coulombic efficiency, relatively good cycling stability, and low cost [19, 20], but the relatively low theoretical specific capacity (372 mAh/g) and poor rate capability restricts its further application [21]. In addition, graphite cannot be used in NIBs, because the Na-C interaction is found to be too weak to contribute to the necessary Coulomb interactions [22]. Therefore, seeking for new anode materials to further improve the performance of LIBs and NIBs is urgently needed [23-25].

2D materials [26, 27] are of special interest as anode materials for LIBs and NIBs because of their high surface area, remarkably high electron mobility and superior mechanical properties [28]. In recent years, many 2D materials have been investigated as anode materials and great success has been obtained [29-44]. Graphene represents the first example of a 2D electrode material for LIBs [45, 46]. Since then, novel 2D materials including $MoS_2$ [47], $VS_2$ [48], silicene [49], phosphorene [50-52], borophene [29, 53], borophane [35], boron phosphide [36], $Mo_2C$ [38], and so on have been investigated as anode materials for LIBs or NIBs. In addition, 2D transition metal carbides or nitrides, called MXenes [54, 55], have also attracted great interest in this field [30-32]. The MXenes can be synthesized by selective etching of A atoms from MAX phases with hydrofluoric acid (HF) at room temperature [54, 55] and the advantage of easy fabrication offers an intrinsic potential for their practical applications [54, 56-59]. Although many 2D materials have been confirmed theoretically as potential electrode materials, the search for LIBs and NIBs with better performance is still necessary and significant. Very recently, Guo et al [60]. have investigated new 2D TMBs such as $Fe_2B_2$ and $Mo_2B_2$ for LIBs. As far as we know, this is the first time that 2D TMBs have been investigated as LIBs. These 2D TMBs belong to orthorhombic system and can be obtained from layered orthorhombic TMBs that possess highly structural similarity to the MAX phases. In addition to these orthogonal structures, there are also a family of layered TMBs which belong to hexagonal crystal system with the formula of $TMB_2$. These layered TMBs contain graphene-like honeycomb boron layers and can be transformed into 2D "sandwich" structures which consist of two boron honeycomb sheets and an intermediate hexagonal plane of TM atom (B-TM-B) [61, 62]. However, because the outermost sheets of MXenes are composed of metallic atoms, we are interested in whether the hexagonal TMBs can also be transformed into 2D structure consisting of an intermediate boron honeycomb and two outer hexagonal planes of TM atoms (TM-B-TM). Besides, it is our great interest to investigate the potential applications of these 2D materials as LIBs and NIBs.

In this work, we first report a series of 2D TMBs including $Sc_2B_2$, $Ti_2B_2$, $V_2B_2$, $Cr_2B_2$, $Y_2B_2$, $Zr_2B_2$, and $Mo_2B_2$. This monolayer $TM_2B_2$ belongs to the space group of P6/mmm with 4 atoms in a hexagonal unit cell, which is identified consisting of an intermediate boron honeycomb sheet sandwiched in between two hexagonal planes of TM atoms. These 2D TMBs can be produced from their bulk phase which is a family of layered TMBs of hexagonal crystal system with the formula of $TMB_2$. Furthermore, we choose $Ti_2B_2$ as the representative and investigate its performance as an anode material for LIBs and NIBs by performing first-principles calculations. Our results illustrate that $Ti_2B_2$ monolayer is a promising electrode material for LIBs and NIBs. This fact is encouraging and indicates that our predicted series of 2D TMBs could be promising candidates for the next-generation portable batteries.

## 2. Computational method

The particle-swarm optimization (PSO) scheme, as implemented in the CALYPSO code [63-65], is employed to search for low energy 2D $Ti_2B_2$ structures. The underlying energy calculations and structure optimizations are performed by using the plane-wave-based density-functional theory (DFT) method as implemented in the Vienna ab-initio Simulation Package (VASP) [66-68]. The computational details of the PSO and DFT methods are illustrated in the Supporting Information. The adsorption energy of lithium and sodium atoms on the $Ti_2B_2$ monolayer is obtained from:

$$E_{ad} = \left(E_{Ti_2B_2M} - E_{Ti_2B_2} - nE_M\right)/n (M = Li, Na) \tag{1}$$

where $E_{Ti_2B_2M}$ and $E_{Ti_2B_2}$ represent the total energies of the metal-ion adsorbed monolayer system and the isolated monolayer, respectively, $E_M$ represents the total energy of per atom for the bulk metal. The charge density difference $\Delta\rho$ are calculated based on the following equation:

$$\Delta\rho = \rho_{Ti_2B_2M} - \rho_{Ti_2B_2} - \rho_M (M = Li, Na) \tag{2}$$

where $\rho_{Ti_2B_2M}$, $\rho_{Ti_2B_2}$, and $\rho_M$ are the total charge of the Li/Na adsorbed system, the $Ti_2B_2$ monolayer, and the Li/Na atom, respectively. The open circuit voltage can be obtained using a well-established approach [69] according to the following formula:

$$V = -\left(E_{Ti_2B_2M} - E_{Ti_2B_2} - nE_M\right)/n (M = Li, Na) \tag{3}$$

To assess the adsorption stability of Li/Na layer on the $Ti_2B_2$ monolayer, the average adsorption energies for each Li/Na layer are calculated according to the following equation:

$$E_{ave} = (E_{Ti_2B_2M_n} - E_{Ti_2B_2M_{(n-1)}} - mE_M)/m (M = Li, Na) \tag{4}$$

where $E_{Ti_2B_2M_n}$ and $E_{Ti_2B_2M_{(n-1)}}$ are the total energies of 2D $Ti_2B_2$ with n and (n-1) adsorbed Li/Na layers, $E_M$ stands for the total energy of per atom for the bulk metal. The number m represents m adsorbed Li/Na atoms in each layer (for a 2×2 supercell on both sides). The theoretical capacity can be obtained from:

$$C_A = czF / M_{Ti_2B_2} \tag{5}$$

where $c$ is the number of adsorbed Li/Na ions, $z$ is the valence number of Li/Na, $F$ is Faraday constant (26801 mAh/mol), and $M_{Ti_2B_2}$ is the molar weight of $Ti_2B_2$.

## 3. Results and Discussion

### 3.1 2D TM$_2$B$_2$ and Bulk TMB$_2$.

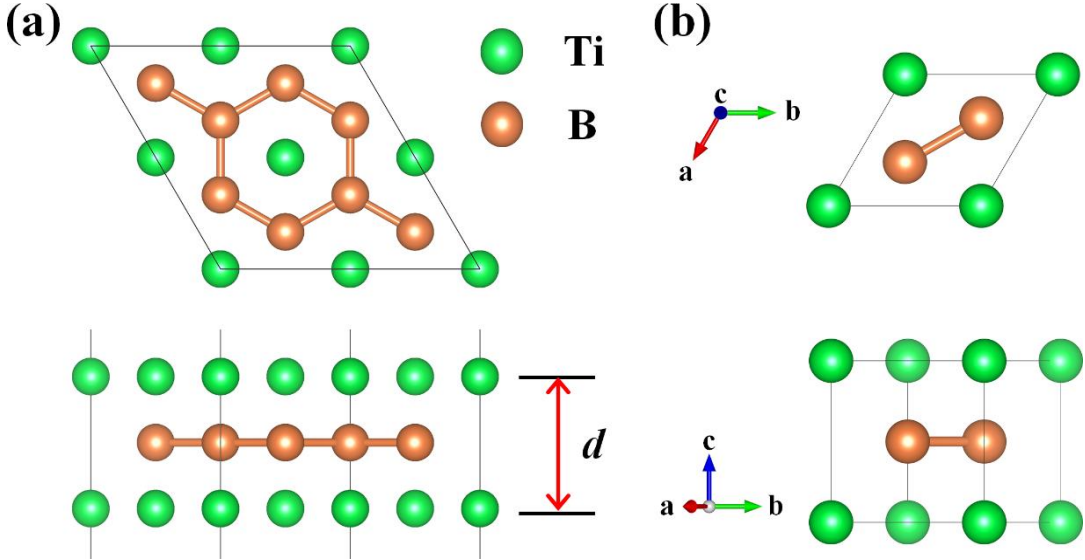

**Fig. 1.** Top and side views of (a) global minimum phase of Ti$_2$B$_2$ monolayer and (b) bulk TiB$_2$. The Ti and B atoms are denoted by green and orange spheres, respectively.

Motivated by the excellent performance of the MXenes in energy storage [70], we investigate TM-B systems characterized by 2D "sandwich" structures. Interestingly, through structural prediction by using the CALYPSO code, we find several 2D Ti$_2$B$_2$ structures. The global minimum structure is shown in Fig. 1a and the other metastable isomers in Fig. S1 in the Supplementary information (SI). The obtained orthorhombic 2D Ti$_2$B$_2$ monolayer (Fig. S1a) possesses the same structure as those of 2D Fe$_2$B$_2$ and Mo$_2$B$_2$ reported by Guo *et al* [60]. According to our calculations, the energy of this orthorhombic 2D Ti$_2$B$_2$ monolayer is higher by 0.067 eV per atom than that of the obtained hexagonal 2D Ti$_2$B$_2$ which is the global minimum structure (Fig. 1a). This hexagonal 2D Ti$_2$B$_2$ consists of an intermediate boron honeycomb sheet sandwiched in between two hexagonal planes of Ti atoms.

Each boron atom bonds to three boron neighbors and the calculated B-B bond length is 1.734 Å. The line charge density distribution along B-B bond is 0.121 e/au$^3$, which is larger than 0.104 e/au$^3$ found for the Si covalent bond [71]. This indicates that the B layers are strongly bonded by B-B covalent bonds. Each Ti atom bonds to six B neighbors and the calculated Ti-B bond length is 2.332 Å. The bulk phase of this 2D structure crystallize in the hexagonal *P6/mmm* space group (AlB$_2$ type No. 191) with *N*=3 in the unit cell stacked along the *c*-axis (see Fig. 1b). The atomic Wyckoff positions of bulk TiB$_2$ are: Ti in 1*a* (0,0,0) and B in 2*d* (1/3,2/3,1/2) [72]. As far as we know, many hexagonal bulk TMB$_2$ structures have been synthesized and all confirmed to possess layered structural properties [73-76]. Thus, we further investigate the viability in other 2D TM$_2$B$_2$ monolayer. By substituting the Ti atom in 2D Ti$_2$B$_2$ by other TM atoms, we find that the hexagonal 2D TM$_2$B$_2$ geometry can be maintained for TM = Sc, V, Cr, Y, Zr, Mo.

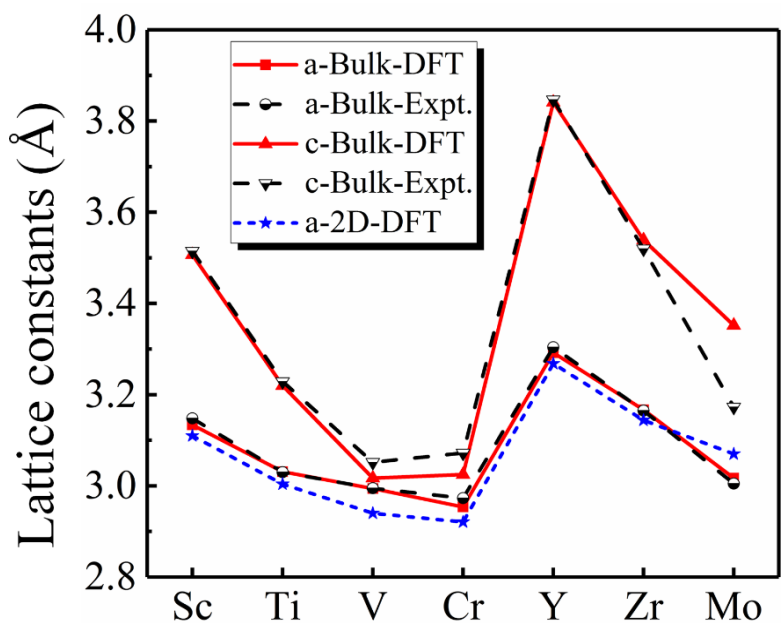

**Fig. 2.** Lattice constants of the bulk TMB$_2$ phase and the 2D TM$_2$B$_2$ structure.

The optimized lattice constants (both *a* and *c*) of these bulk TMB$_2$ are shown in Fig. 2 and Table S1 in the SI, consistent well with the experimental results reported earlier [72-75, 77]. The optimized lattice constant of the 2D Ti$_2$B$_2$ is *a* = *b* = 3.004 Å, comparable to that of the bulk TiB$_2$ (3.031 Å) [72]. The optimized thickness of the 2D Ti$_2$B$_2$ is *d* = 3.117 Å, also comparable to the lattice constant *c* of the bulk TiB$_2$ (3.220

Å) [72]. As shown in Fig. 2, the optimized lattice constants (*a*) of other six 2D TM$_2$B$_2$ are all consistent with those of their corresponding bulk structures. This indicates that these 2D structures are easily obtained from their corresponding bulk phases.

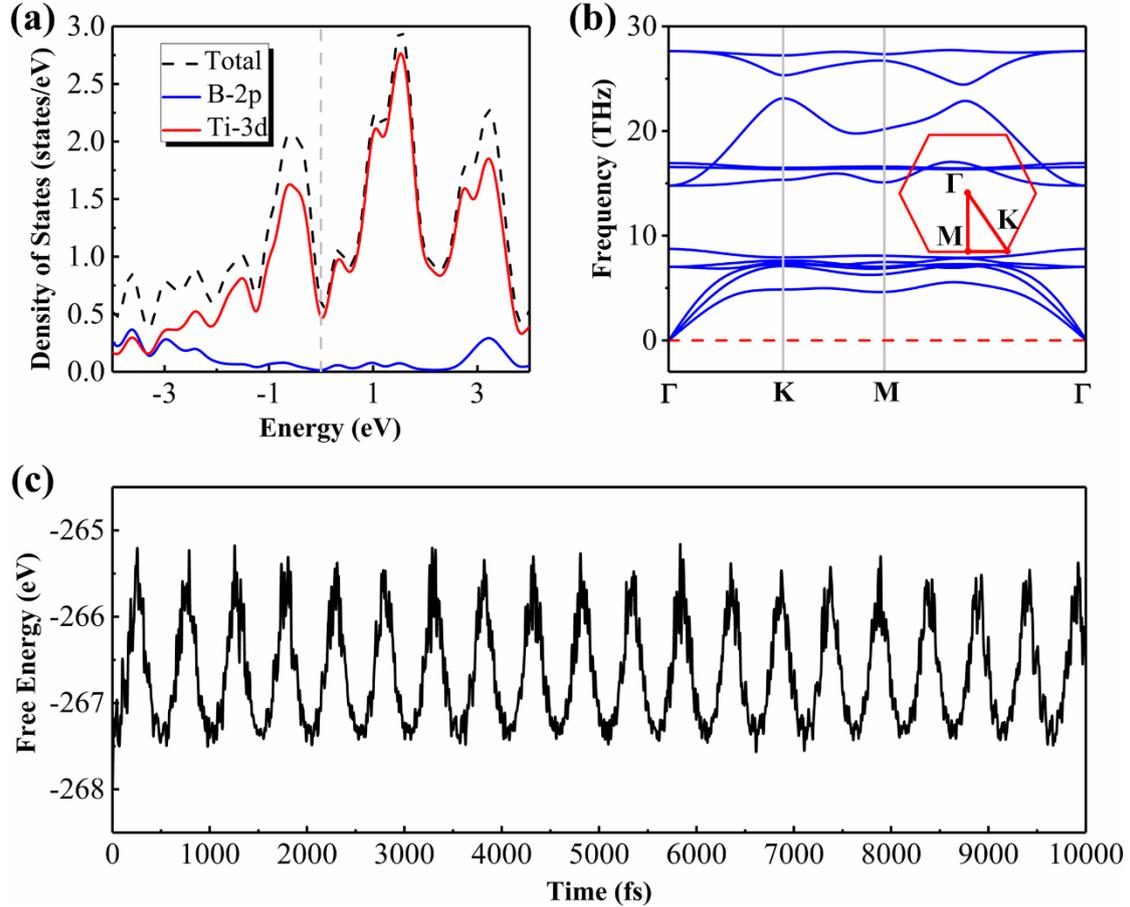

**Fig. 3.** (a) Total and partial density of states (TDOS and PDOS) of the 2D Ti$_2$B$_2$. The dash black line represents the TDOS while the solid red and solid blue lines stand for the Ti-3*d* and B-2*p* PDOS, respectively. The Fermi energy level is set at zero. (b) Phonon dispersion curves of the 2D Ti$_2$B$_2$. (c) Variation of the free energy in the AIMD simulations at 300 K during the time scale of 10 ps.

As shown in Fig. 3a, the 2D Ti$_2$B$_2$ exhibits metallic property contributed mainly by the Ti-3*d* orbitals. The contribution from the B-2*p* is ignorable. The TDOS and PDOS of 2D Sc$_2$B$_2$, V$_2$B$_2$, Cr$_2$B$_2$, Y$_2$B$_2$, Zr$_2$B$_2$, and Mo$_2$B$_2$ (shown in Fig. S2 in the SI) indicate that these six 2D materials, like 2D Ti$_2$B$_2$, present good conductivity. Owing to the outstanding electronic conductivity of these 2D TM$_2$B$_2$, they merit potential application as anode materials for LIBs and NIBs.

The stability of anode materials is critical important due to hundreds of charge-discharge cycles in practical applications. To examine the dynamic stability of 2D $Ti_2B_2$, we calculate its phonon dispersion curves. As shown in Fig. 3b, the absence of imaginary modes in the whole BZ confirms that 2D $Ti_2B_2$ is dynamically stable. The highest frequency of 2D $Ti_2B_2$ reaches up to 27.725 THz (924 cm$^{-1}$), which is higher than those of $FeB_2$ (854 cm$^{-1}$) [78], TiC (810 cm$^{-1}$) [79], $Cu_2Si$ (420 cm$^{-1}$) [80], and $MoS_2$ (473 cm$^{-1}$) [81] monolayers. The high value of frequencies in the phonon dispersion also indicates the stability of this 2D material. The phonon spectra of other six 2D $TM_2B_2$ are also calculated and shown in Fig. S3 in the SI. We find that these 2D $TM_2B_2$ are also dynamically stable. To check the thermal stability, we carry out AIMD simulations for the 2D $Ti_2B_2$ at temperatures of 300, 600, 1200, 1800, 2400, and 3000 K, respectively. At 300 K, the average value of free energy remains almost constant during the whole simulation (see Fig. 3c), confirming that the 2D $Ti_2B_2$ is thermally stable at room temperature. Even increasing the temperature up to 2400 K, the original geometry remains intact with only slight in-plane and out-of-plane deformations (see Fig. S4). The free energies fluctuate around constant values during the AIMD simulation below and at 2400 K (see Fig. S5). These results confirm that the 2D $Ti_2B_2$ is stable upon heating, which, supplies safeguard for practical applications.

**3.2 Adsorption of Li and Na on the 2D $Ti_2B_2$ monolayer.**

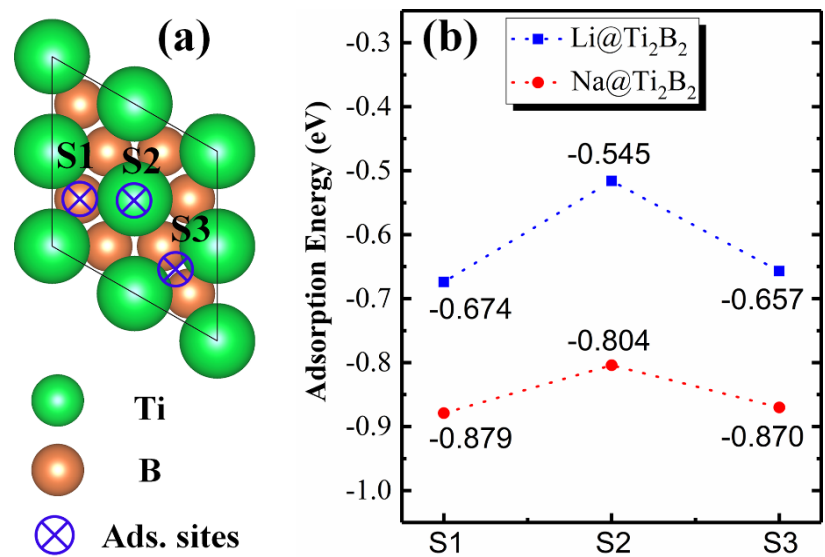

**Fig. 4.** (a) High symmetry adsorption sites and (b) adsorption energies for Li/Na on the surface of 2D $Ti_2B_2$. The Ti and B atoms are denoted by green and orange spheres, respectively.

The intrinsic metallicity and good stability of the 2D $Ti_2B_2$ make it a promising anode material for LIBs and NIBs. Next, we will further check its ability of adsorbing Li/Na atoms. We first investigate the adsorption behaviors of a single Li/Na atom on the surface of the 2D $Ti_2B_2$ by considering three high symmetry adsorption sites S1, S2, and S3 (see Fig. 4a). The adsorption energies of Li/Na atom on these sites are shown in Fig. 4b. We find that the Na-adsorbed configurations are more stable than the Li-adsorbed ones and the S1 is always the most stable adsorption site for both Li and Na atoms. Interestingly, the adsorption energies on S3 are very close to those on S1, suggesting that the Li/Na atoms are easy to diffuse along the S1-S3-S1 direction, i.e. the B-B bonding direction. The adsorption energies on S2 are smaller by 0.06-0.13 eV than those on S1 and S3. This indicates that the Li/Na atom is not possible to appear on top of the Ti atoms. To better understand the adsorption properties of Li/Na on the $Ti_2B_2$ surface, we calculate the difference charge density and present them in Fig. 5 (Li-adsorption) and Fig. S6 (Na-adsorption). It is clear that the electrons tend to accumulate in between Li/Na and its neighbor Ti atoms, thus, result in the Li/Na-Ti bonding. This chemical bonding between Li/Na atoms and 2D $Ti_2B_2$ are favorable to prevent the forming of Li/Na cluster and improve the safety and application availability for Li/Na storage in LIBs and NIBs.

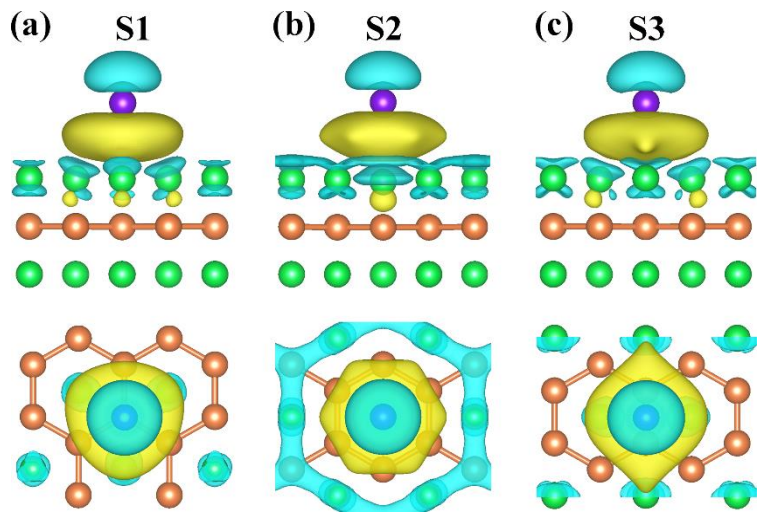

**Fig. 5.** The charge density difference plots for Li adsorption on the (a) S1, (b) S2, and (c) S3 sites of 2D $Ti_2B_2$ monolayer. The yellow and cyan areas represent electron gains and loses. The Ti, B, and Li atoms are denoted by green, orange, and violet spheres, respectively.

**3.3 Theoretical specific capacity and average open-circuit voltage.**

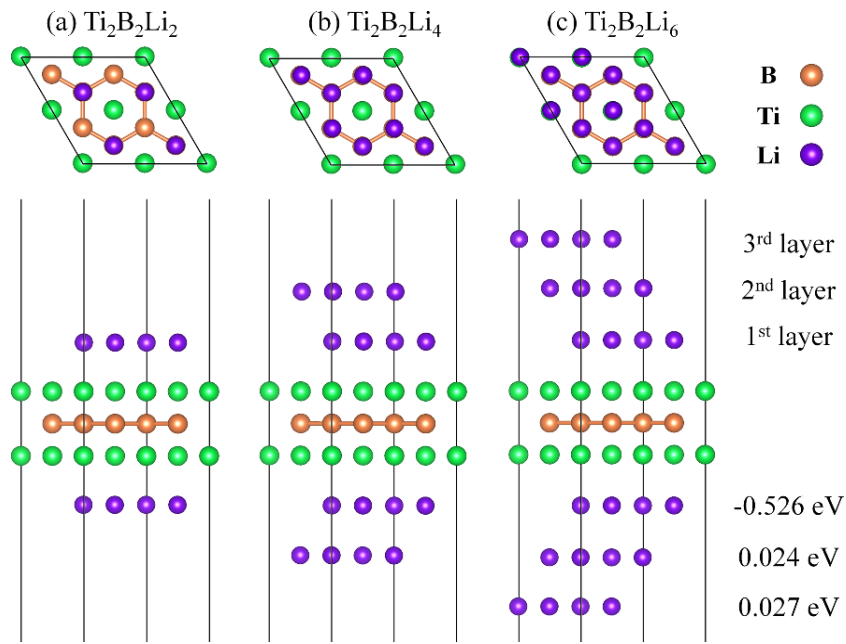

**Fig. 6.** Top and side views of the structures of (a) $Ti_2B_2Li_2$, (b) $Ti_2B_2Li_4$, and (c) $Ti_2B_2Li_6$ with layers of Li ions adsorbed on the surface of 2D $Ti_2B_2$ monolayer. The Ti, B, and Li atoms are denoted by green, orange, and violet spheres, respectively.

For practical application, it is of great significance to investigate the storage capacity of the batteries for the electrode materials. Thus, the average adsorption energies are calculated to investigate the storage capacity of Li/Na on the $Ti_2B_2$ monolayer. The adsorptions of one, two, and three layers of Li on both sides of $Ti_2B_2$ monolayer are investigated using 2×2 supercells and can be labeled as $Ti_2B_2Li_x$ with $x$=2, 4 and 6, respectively. For the one-layer adsorption, shown in Fig. 6a, all Li atoms are absorbed at the S1 sites. In condition of saturation, there are eight Li atoms that can be adsorbed on both sides of the $Ti_2B_2$ monolayer. The average adsorption energy of Li atoms is -0.526 eV, which indicates that Li adatoms can be adsorbed stably without clustering. After the first layer being adsorbed fully, the subsequently added

Li atoms will form the second adsorbed layer. The Li atoms are also adsorbed at the S1 sites (see Fig. 6b). This can be regard as the two-layer adsorption and there are 16 Li atoms being adsorbed in condition of full adsorption. For the three-layer adsorption, the third-layer Li atoms are adsorbed at the S2 sites (Fig. 6c). In condition of full adsorption, there are 24 Li atoms being adsorbed. The adsorption energies for one-layer, two-layer, and three-layer adsorption are -4.208, -4.019, and -3.800 eV, respectively. However, according to Eq. (4), the calculated average adsorption energies for the second and third layers of Li atoms are 0.024 and 0.027 eV, respectively. The positive values of these adsorption energies mean that the adsorptions of two and three layers of Li are not stable in energy. The above results reveal that the 2×2 supercell of $Ti_2B_2$ monolayer can accumulate 8 Li atoms at most, corresponding to $Ti_2B_2Li_2$ with symmetric configuration of adatoms on both sides of $Ti_2B_2$ monolayer. According to these results and Eqs (3) and (5), the corresponding theoretical specific capacity and average open-circuit voltage of the 2D $Ti_2B_2$ as electrode of LIBs are ~456 mA h $g^{-1}$ and 0.526 eV, respectively.

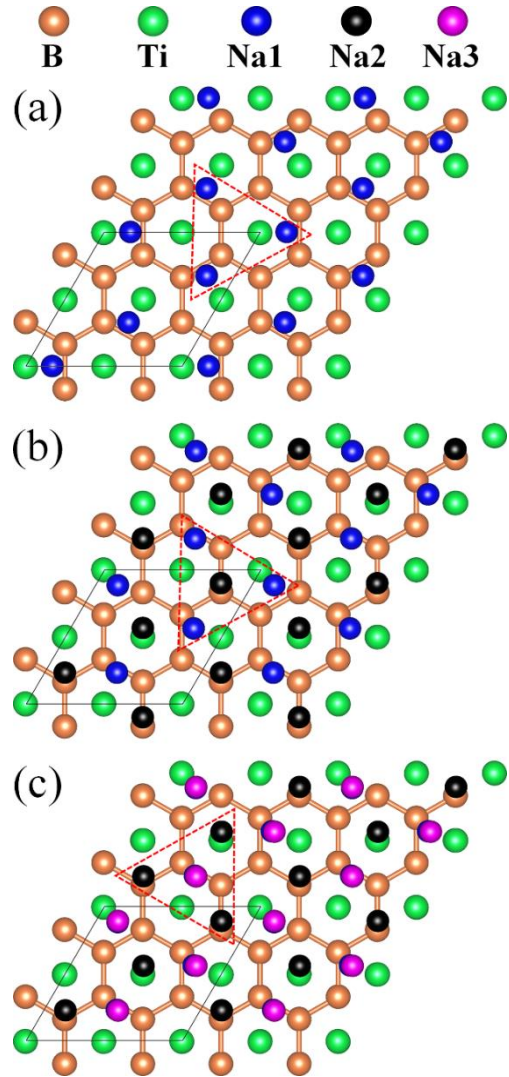

**Fig. 7.** Top views of (a) one-layer, (b) two-layers, and (c) three-layers adsorptions of Na ions on the 2D $Ti_2B_2$ monolayer. Na1, Na2, and Na3 represent the Na ions adsorbed in the first, the second, and the third layer, respectively. The black lines stand for the range of a 2×2 supercell.

Next, we turn our focus on the case of Na-adsorbed $Ti_2B_2$ system. The one-layer, two-layers, and three-layers adsorptions of Na on both sides of 2D $Ti_2B_2$ are calculated and the corresponding schematic pictures are presented in Fig. 7. For clarity, only Na atoms on single side of the 2D $Ti_2B_2$ are shown in this figure. The first layer can accommodate at most six Na atoms on both sides of the $Ti_2B_2$ monolayer, three Na atoms on each side. This number is less than that of Li adsorption. Since the atomic radius of Na is larger than that of Li, this fact is understandable. As shown in Fig. 7a, the three Na atoms form an equilateral triangle and the adsorption sites are not at the S1. When the first layer is adsorbed fully, the additionally added Na atoms will form the second layer at the sites above the center of the equilateral triangles (Fig. 7b). The adsorption sites for the third layer are above the center of the equilateral triangles of the second layer (Fig. 7c). The second and the third layers can also accommodate as many as six Na atoms on both sides. The average adsorption energies are all negative for the first, the second, and the third layers (-0.502, -0.007, and -0.009 eV, respectively), indicating that Na atoms can be adsorbed stably without clustering. Furthermore, the adsorption of four layers of Na on the $Ti_2B_2$ monolayer is not energetically stable according to our calculation. The above results reveal that the 2×2 supercell of $Ti_2B_2$ monolayer can accommodate 18 Na adatoms, corresponding to $Ti_2B_2Na_{4.5}$. Therefore, the calculated theoretical specific capacity of the 2D $Ti_2B_2$ as electrode of NIBs is ~1027 mA h $g^{-1}$. The average open-circuit voltage decreases from 0.502, via 0.255, to 0.173 eV with increasing of the adsorbed Na atoms from 6, via 12, to 18.

We also estimate the varying of lattice constant and thickness of the $Ti_2B_2$ monolayer after the adsorption of layered Li/Na atoms. The lattice constant increases from 6.008 ($Ti_2B_2$) to 6.047 Å ($Ti_2B_2Li_2$) for Li adsorption (about 0.65% tensile strain), and slightly increases from 6.008 ($Ti_2B_2$), 6.040 ($Ti_2B_2Na_{1.5}$), and 6.054 ($Ti_2B_2Na_3$) to 6.070 Å ($Ti_2B_2Na_{4.5}$) for Na adsorption (about 1.03% tensile strain). The calculated thickness of the $Ti_2B_2$ monolayer for the one-layer adsorbed Li structure ($Ti_2B_2Li_2$) is 3.128 Å, which is larger by 0.29% than that of the bare monolayer (3.119 Å). However, the thicknesses of the $Ti_2B_2$ monolayer for the one-layer ($Ti_2B_2Na_{1.5}$), two-layer ($Ti_2B_2Na_3$), and three-layer ($Ti_2B_2Na_{4.5}$) adsorptions of Na are 3.108, 3.106, and 3.110 Å, respectively. These values are smaller by 0.35%, 0.42%, and 0.29% than the bare one, respectively. In view of these results, the $Ti_2B_2$ monolayer is robust to be used as anode material in both LIBs and NIBs.

## 3.4 Electronic structures of the 2D Ti$_2$B$_2$ monolayer with adsorbed Li/Na atoms.

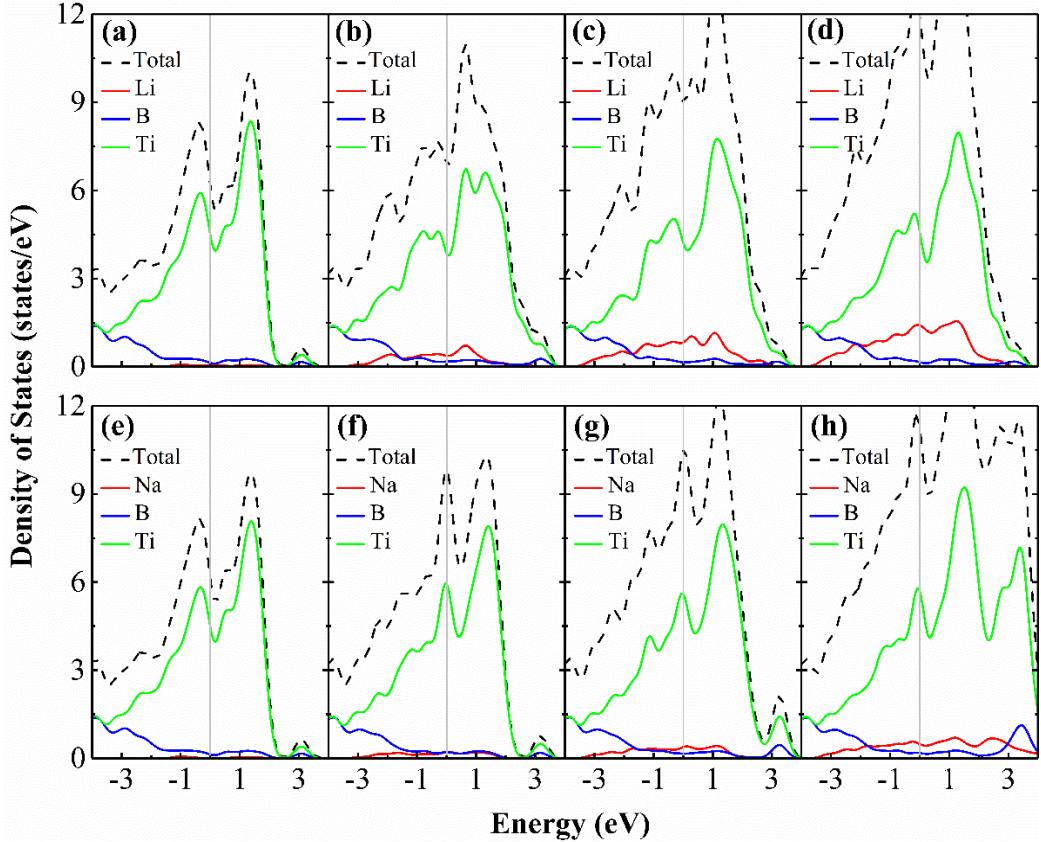

**Fig. 8.** TDOS and PDOS of the 2D Ti$_2$B$_2$ monolayer adsorbed with (a) one Li atom, (b) one-layer Li atoms, (c) two-layer Li atoms, (d) three-layer Li atoms, (e) one Na atom, (f) one-layer Na atoms, (g) two-layer Na atoms, and (h) three-layer Na atoms, respectively. The Fermi energy levels are set as zero.

It is well known that the electronic conductivity is closely associated with the rate capability of the anode materials in LIBs and NIBs. Unlike many other 2D materials that exhibit semi-conducting or semi-metallic characteristics [82, 83], our previous analyses have demonstrated that the pristine Ti$_2$B$_2$ monolayer exhibits intrinsic metallic behavior, which means a remarkable battery performance for the sheet. In order to gain insight into the electronic conductivity of the charge-discharge process, the evolutions of DOS for Ti$_2$B$_2$ monolayer adsorbed with Li/Na atoms are calculated (Fig. 8). Fig. 8a shows the DOS of single Li adsorbed on the Ti$_2$B$_2$ monolayer. Compared to the DOS of the pristine Ti$_2$B$_2$ monolayer (Fig. 3a), we find that the orbital contributions around the Fermi level mainly come from the sheet with negligible components from the Li atoms, endowing the system with metallic characteristic. To be mentioned that $Ti_2B_2$ monolayer is easy to be terminated with multiple Li atoms, hence the modulation effect of surface atomic number on electronic structures has also been evaluated in the following sequence: one-layer (Fig. 8b), two-layer (Fig. 8c), and three-layer (Fig. 8d) adsorbed Li configurations. With the Li concentration increasing during the charging process, the orbital contributions from Li atoms near the Fermi level increase with the components of the monolayer showing no big fluctuations. Meanwhile, the Fermi level is getting away from the valley bottom of the DOS with the increasing number of Li atoms, indicating the growing instability of the system. These simulations explicitly demonstrate that the system retains the metallic nature during Li atomic adsorption process. Similar evolutions of DOS, for the host $Ti_2B_2$ monolayer with one-layer, two-layer, and three-layer Na adsorbed, are also found in Fig. 8. These robust electronic properties of $Ti_2B_2$ monolayer greatly expand its potential applications in anode materials and motivate subsequent experimental studies.

**3.5 Li/Na diffusion on 2D $Ti_2B_2$ monolayer.**

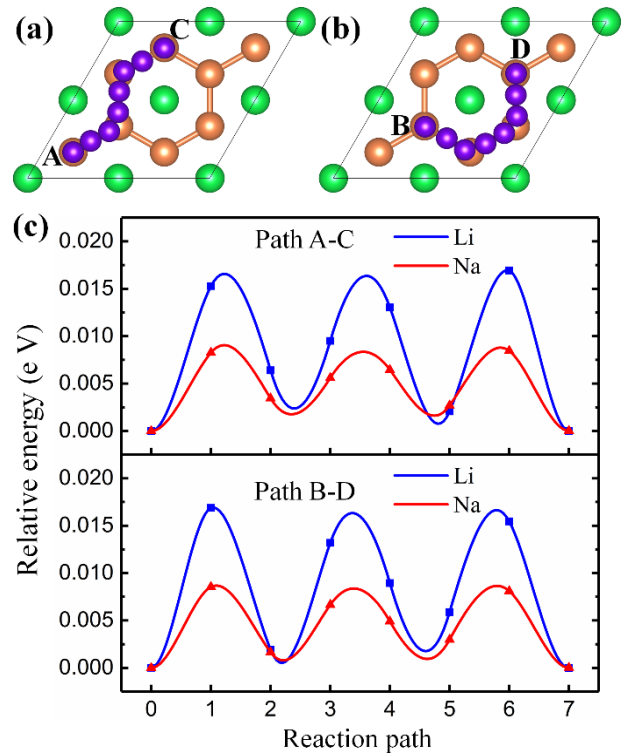

**Fig. 9.** Top view of two pathways, (a) Path A-C and (b) Path B-D, for Li/Na diffusion on the 2D $Ti_2B_2$ monolayer. The Ti, B, and Li/Na atoms are denoted by green, orange, and violet spheres, respectively. (c) Energy profiles of the corresponding Li and Na diffusion pathways.

In addition to the high specific capacity, a good charge−discharge rate capability is of great significance for a promising anode material for LIBs and NIBs. The diffusion barrier of Li/Na is a key factor that determines the charge-discharge rate. In this part, we employ the CI-NEB method to investigate the diffusion barriers for Li and Na atoms on the surface of $Ti_2B_2$ monolayer. As Li and Na atoms preferentially occupy the S1 site, two possible pathways starting from one S1 site to another S1 site are considered. The first pathway is from A to C (Fig. 9a) and the second from B to D (Fig. 9b). Both these pathways are connected by four S1 and three S3 sites and can be viewed as S1-S3-S1-S3-S1-S3-S1. As illustrated in Fig. 9c, the energy barriers for Li diffusion along the A-C and B-D pathways are both 0.017 eV while that for Na diffusion are both 0.008 eV. The ultralow diffusion energy barriers indicate that Li and Na atoms can diffuse extremely easy on the $Ti_2B_2$ monolayer. Furthermore, due to the smaller diffusion energy barrier of Na than that of Li, the Na ions are easier to diffuse on the $Ti_2B_2$ monolayer than Li ions. The larger rate capacity for Na than Li is also found in other 2D materials, such as $B_2H_2$ [35] and $Mo_2C$ [38]. Here, the low diffusion energy barriers are due to the small difference of the Li/Na adsorption energies between the S1 and S3 sites. Since that the diffusion energy barriers for A-C and B-D pathways are the same, the two pathways are both possible routes for Li/Na ion diffusion, which, would beneficial to increase the charge-discharge rate of LIBs and NIBs.

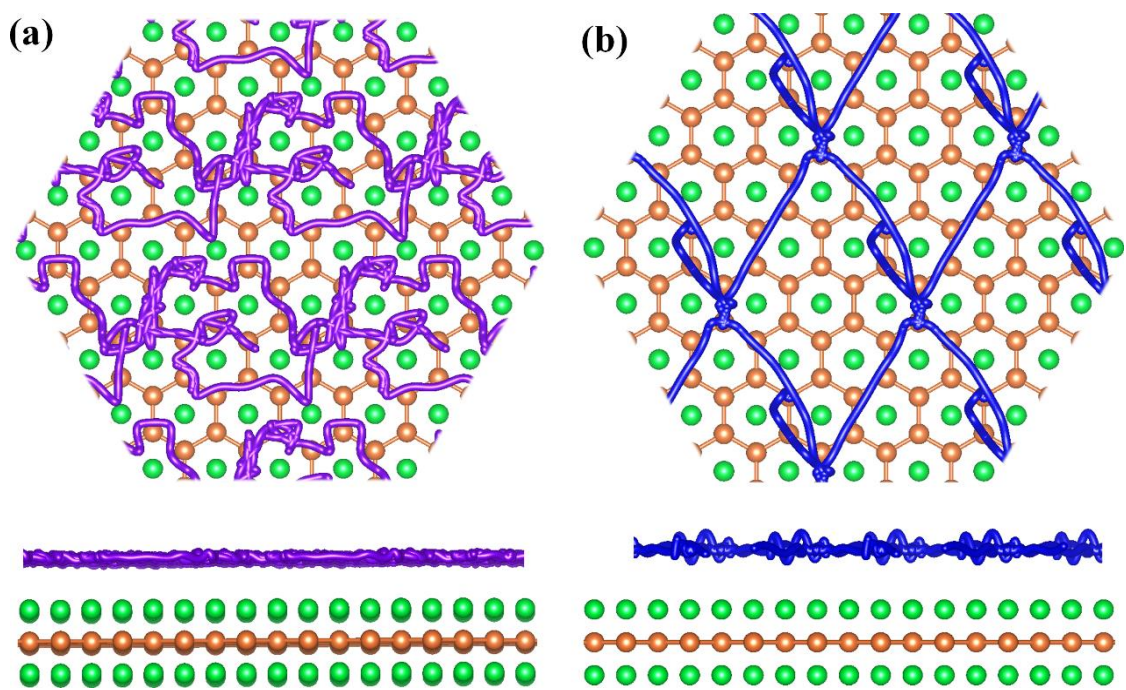

**Fig. 10.** Top and side views of the calculated (a) Li and (b) Na trajectories during the 10 ps AIMD simulations. The Ti, B, Li, and Na atoms are denoted by green, orange, violet, and blue spheres, respectively.

To further investigate the diffusion kinetics of Li and Na atoms on the $Ti_2B_2$ monolayer, the AIMD simulations are performed at 300 K lasting for 10 ps. The Li and Na diffusion trajectories during the 10 ps simulation, shown in Fig. 10a and 10b, respectively, give direct visualizations of Li and Na diffusion on the $Ti_2B_2$ monolayer. We see that the Li atom can travel through the entire region almost freely along the A-C or B-D pathways, although the diffusion path is not that straight. Interestingly, the Na atom can diffuse through the entire region in almost straight lines along the A-C pathway. Similar behavior has been observed on borophene for Li diffusion [29]. The above AIMD simulation results explicitly exhibit an ultra-fast Li/Na diffusion on the 2D $Ti_2B_2$ monolayer, giving a proof that this 2D material is a promising candidate for the high rate LIBs and NIBs electrode materials.

**3.6 Comparison with other anode materials.**

In order to evaluate the superiority of our studied $Ti_2B_2$ monolayer as an electrode material for LIBs and NIBs, we compare the theoretical specific capacity and diffusion barrier of it with other widely investigated anode materials [8, 29, 31, 33-35, 38, 47-53, 60, 84-92] in Table 1 and Table 2. For Li storage, the theoretical specific capacity of the $Ti_2B_2$ monolayer is 456 mAhg$^{-1}$, which is greater than that of the well-established graphite [8, 84, 85], phosphorene [50, 51], $Ti_3C_2$ [31], or $Li_4Ti_5O_{12}$ [86], and comparable to that of the $Mo_2C$ [38], $VS_2$ [48], $B_2H_2$ [35], and planar $TiB_4$ [92]. Compared to the orthorhombic 2D MBenes [60], the theoretical specific capacity of the hexagonal 2D $Ti_2B_2$ is in between those of the 2D $Mo_2B_2$ and the 2D $Fe_2B_2$ [60]. Although the theoretical specific capacity is lower than that of silicon [87], silicene [49], and Sn [86, 88], the energy barrier (0.017 eV) for Li diffusion is the lowest one among the listed materials in Table 1 besides borophene [29]. This indicates that the 2D $Ti_2B_2$ possesses a very high rate capability for Li diffusion.

In the case of Na storage and diffusion, as shown in Table 2, the $Ti_2B_2$ monolayer has the second largest theoretical specific capacity. The value of 1027 mAhg$^{-1}$ is larger than that of phosphorene [89] and boron-doped graphene [90] and close to the highest theoretical specific capacity we found in literature for $Ca_2N$ [34]. When compared with other 2D materials, the theoretical specific capacity of $Ti_2B_2$ is about 2 to 3 times of that reported in $Ti_3C_2$ [33], borophene [53], $B_2H_2$ [35], and $Sr_2N$ [34], and is more than 7 times of that reported in $MoS_2$ [47] and $Mo_2C$ [38]. In addition, the diffusion energy barrier (0.008 eV) of Na ion on $Ti_2B_2$ monolayer is much lower than that of any other typical promising anode materials, shown in Table 2, indicating a very high rate capability.

In the end, we also compare the properties as an anode material for 2D $Ti_2B_2$ with the hexagonal 2D $TiB_4$ whose atomic stacking is B-Ti-B (Fig. S7a). The calculated energy of this 2D $TiB_4$ (Fig. S7a) is lower by 0.198 eV per atom than the most stable planar 2D $TiB_4$ (Fig. S7b) [92], indicating the stability of this hexagonal 2D $TiB_4$. The most stable adsorption sites and the corresponding adsorption energies are presented in Fig. S8. The 2×2 supercell of the hexagonal 2D $TiB_4$ can accommodate 8 Li or Na atoms (Fig. S9), corresponding to $TiB_4Li_2$ and $TiB_4Na_2$, respectively. The theoretical specific capacity of this hexagonal 2D $TiB_4$ are both 588 mAhg$^{-1}$ for Li and Na storage, indicating a higher Li storage but a lower Na storage than those of the 2D

Ti$_2$B$_2$. However, the diffusion energy barriers of Li (0.712 eV) and Na (0.377 eV) on the hexagonal 2D TiB$_4$ are much larger than those on the 2D Ti$_2$B$_2$ (Fig. S10), indicating that the 2D Ti$_2$B$_2$ is more suitable to be an electrode material than the 2D TiB$_4$.

**Table 1.** Summary of theoretical specific capacities (mAhg$^{-1}$) and diffusion barriers (meV) of some widely investigated promising anode materials for LIBs.

| Species | Theoretical specific capacity | Diffusion barrier | References |
|---|---|---|---|
| Ti$_2$B$_2$ | 456 | 17 | This work |
| Graphite | 372 | 450~1200 | [8, 84, 85] |
| Phosphorene | 433 | 80 | [50, 51] |
| Silicon | 4200 | 580 | [87] |
| Silicene | 954 | 230 | [49] |
| Sn | 994 | 390 | [86, 88] |
| 1T-Ti$_3$C$_2$ | 320 | 70 | [31] |
| 1H-Mo$_2$C | 526 | 35 | [38] |
| VS$_2$ | 466 | 220 | [48] |
| Borophene | 1860 | 2.6 | [29] |
| B$_2$H$_2$ | 504 | 210 | [35] |
| Li$_4$Ti$_5$O$_{12}$ | 175 | 300 | [86, 91] |
| Orthorhombic Fe$_2$B$_2$ | 665 | 240 | [60] |
| Orthorhombic Mo$_2$B$_2$ | 444 | 270 | [60] |
| Planar TiB$_4$ | 588 | 180 | [92] |

**Table 2.** Summary of theoretical specific capacities (mAhg$^{-1}$) and diffusion barriers (meV) of some widely investigated promising anode materials for NIBs.

| Species | Theoretical specific capacity | Diffusion barrier | References |
|---|---|---|---|
| Ti$_2$B$_2$ | 1027 | 8 | This work |
| Phosphorene | 865 | 40 | [52] |
| graphene |  | 130 | [89] |
| Boron-doped graphene | 762 | 160-220 | [90] |
| 1T-MoS$_2$ | 146 | 280 | [47] |
| 1T-Ti$_3$C$_2$ | 352 | 96 | [33] |
| 1H-Mo$_2$C | 132 | 15 | [38] |
| Borophene | 496/596 | 300/12 | [53] |
| B$_2$H$_2$ | 504 | 90 | [35] |
| Ca$_2$N | 1138 | 80 | [34] |
| Sr$_2$N | 283 | 16 | [34] |

## 4. Conclusion

Searching for suitable electrode materials with good performance is urgently needed for energy storage. In this work, we firstly reported a series of hexagonal 2D transition metal borides (TMBs) including Sc$_2$B$_2$, Ti$_2$B$_2$, V$_2$B$_2$, Cr$_2$B$_2$, Y$_2$B$_2$, Zr$_2$B$_2$, and Mo$_2$B$_2$ by using first-principles method and crystal structure prediction techniques. Through the calculated phonon spectra and DOS, we confirmed that these TMBs possess great stability and excellent electronic conductivity.

Then, we investigated the potential of 2D Ti$_2$B$_2$ monolayer as the anode material for LIBs and NIBs on the basis of first-principles simulations. The Ti$_2$B$_2$ monolayer shows negative adsorption energies for Li and Na of -0.674 and -0.879 eV, respectively, on the most stable S1 site. The calculated theoretical specific capacity values for Li on Ti$_2$B$_2$ monolayer is 456 mAhg$^{-1}$, which is larger than the well-established graphite, phosphorene, and Ti$_3$C$_2$. Interestingly, the theoretical specific capacity for Na is investigated to be 1027 mAhg$^{-1}$, which is much larger than most other 2D materials. More excitingly, we found that the $Ti_2B_2$ monolayer shows very high Li and Na diffusivity with ultralow energy barrier of 0.017 and 0.008 eV, respectively, indicating an excellent charge-discharge capability for Li and Na. Moreover, the $Ti_2B_2$ monolayer was found to exhibit metallic nature after the adsorption of Li and Na ions at any concentrations, indicating that the $Ti_2B_2$ monolayer possesses an excellent electronic conductivity. Finally, the $Ti_2B_2$ monolayer just has a small change in volume after the adsorption of one, two, and three layers of Li and Na ions, indicating the robustness of this material.

Considering the above advantages, it is expected that the $Ti_2B_2$ monolayer and its alike materials would be used as anode materials for LIBs and NIBs with high specific capacity and fast diffusivity.

**Corresponding Author**

*E-mail: wangbt@ihep.ac.cn

**Notes**

The authors declare no competing financial interest.

**Acknowledgments**

F.W.W. and J.R.Z. acknowledge financial support from National Natural Science Foundation of China under Grants No. 11675255, No. 11634008, and No. 11675195. J.R.Z. acknowledges financial support from The National Key Basic Research Program under Grants No. 2017YFA0403700. The calculations were performed at Supercomputer Centre in China Spallation Neutron Source.

**Supplementary information**

Supplementary information associated with this article can be found in the online version at:

(I) Computational method, (II) metastable isomers of 2D $Ti_2B_2$, (III) lattice constants (Å) of the bulk $TMB_2$ phases and the 2D $TM_2B_2$ structures, (IV) DOS and phonon dispersion curves of the other six 2D TM$_2$B$_2$, (V) snapshots and variation of the free energy for 2D Ti$_2$B$_2$ in the AIMD simulations from 300 to 3000 K, (VI) charge density difference plots for Na adsorption, (VII) 2D TiB$_4$ structures and the adsorption and diffusion of Li/Na on the hexagonal TiB$_4$ monolayer.

**Graphical Abstract:**

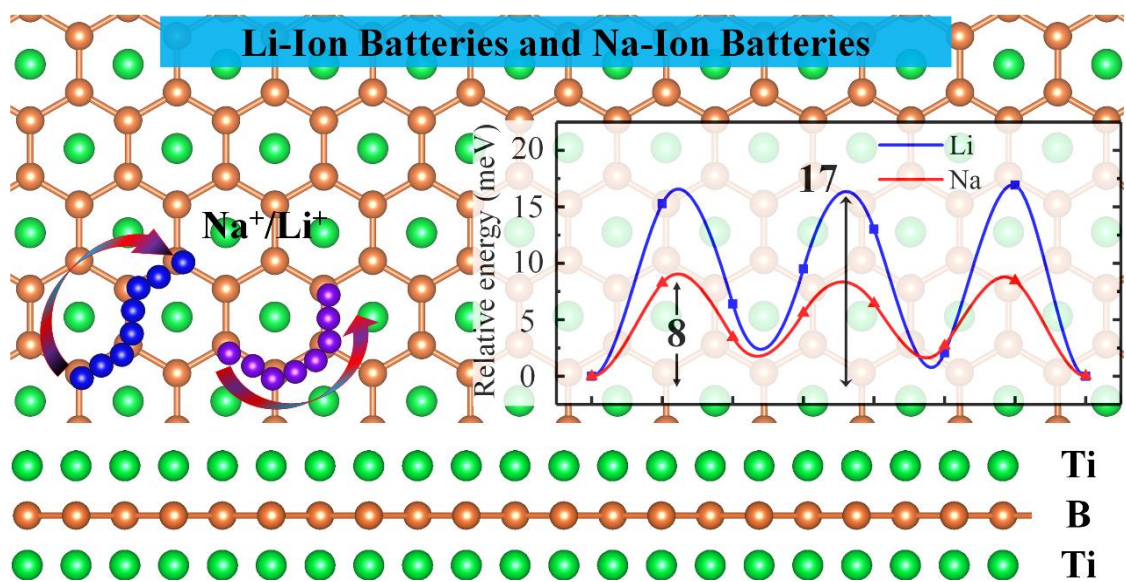